# A collagen-based theranostic wound dressing with visual, long-lasting infection detection capability

Charles Brooker,[1,2] Giuseppe Tronci[1,2*]

[1] Clothworkers' Centre for Textile Materials Innovation for Healthcare (CCTMIH), School of Design, University of Leeds, Leeds, LS2 9JT, United Kingdom

[2] School of Dentistry, St. James's University Hospital, University of Leeds, Leeds, LS9 7TF, United Kingdom

[*] Corresponding author: g.tronci@leeds.ac.uk

## Abstract

Continuous wound monitoring is one strategy to minimise infection severity and inform prompt variations in therapeutic care following infection diagnosis. However, integration of this functionality in therapeutic wound dressings is still challenging. We hypothesised that a theranostic dressing could be realised by integrating a collagen-based wound contact layer with previously demonstrated wound healing capability, and a halochromic dye, i.e. bromothymol blue (BTB), undergoing colour change following infection-associated pH changes (pH: 5-6 → >7). Two different BTB integration strategies, i.e. electrospinning and drop-casting, were pursued to introduce long-lasting visual infection detection capability through retention of BTB within the dressing. Both systems had an average BTB loading efficiency of 99 wt.% and displayed a colour change within one minute of contact with simulated wound fluid. Drop-cast samples retained up to 85 wt.% of BTB after 96 hours in a near-infected wound environment, in contrast to the fibre-bearing prototypes, which released over 80 wt.% of BTB over the same time period. An increase in collagen denaturation temperature (DSC) and red shifts (ATR-FTIR) suggests the formation of secondary interactions between the collagen-based hydrogel and the BTB, which are attributed to count for the long-lasting dye confinement and durable dressing colour change. Given the high L929 fibroblast viability in drop-cast sample extracts (92%, 7 days), the presented multiscale design is simple, cell- and regulatory-friendly, and compliant with industrial scale-up. This design, therefore, offers a new platform for the development of theranostic dressings enabling accelerated wound healing and prompt infection diagnosis.

## 1 Introduction

Due to their delayed healing capability [1], chronic wounds can take several years, and even decades to heal [2], triggering risks of infection, gangrene, haemorrhage, and amputation. [3] In the United Kingdom, chronic wounds cost the National Health Service (NHS) £5bn per annum, due to their delayed healing, the need for prolonged hospitalisation time and clinical complications, such as recurrent infections. [4] Bacterial resistance to antibiotics is therefore a fundamental challenge in the management and treatment of chronic wounds. [5,6]

Wound dressings are routinely used in clinical practice as a cost-effective route to support wound healing and manage infection. [7] A range of therapeutic modalities has been integrated into currently available commercial products, including the encapsulation of silver nanoparticles or the loading of natural antimicrobials, such as honey. [8] At the same time, complications due to bacterial resistance [9] as well as the lack of prompt infection diagnosis and clinical evidence make current dressings unable to effectively accelerate wound healing, reduce hospitalisation and save healthcare costs.

A promising strategy to minimise infection severity is through continuous wound monitoring aiming to inform prompt variations in therapeutic care following rapid infection diagnosis. [10] Specifically, the development of user-friendly infection-monitoring devices is expected to enable patients to spend more time in their own homes, minimising their need to visit the hospital, and reducing the healthcare burden. [11,12] This trend in the shift of healthcare services away from hospitals and clinics and into the local community and patients' homes would be particularly useful in the case of pandemic outbreaks, as observed during the COVID-19 pandemic. [13,14]

Compared to traditional wound dressings, which had an almost sole focus on avoiding strikethrough and protecting the wound from contamination [15], growing research has been carried out aiming to integrate state-of-the-art wound dressings with advanced wound management functionalities. Examples of these functionalities include shape memory capability [16], anti-inflammatory character [17,18], drug delivery [19,20], and sensing capabilities. [21,22] These modern, "smart" dressings with sensing capabilities have been developed to detect in real-time various physiological and biochemical changes at the wound site, monitoring the wound conditions and detecting infections. [23] There are a number of biomarkers which can be used for wound monitoring and for infection detection, which include glucose [24], lactate [25], oxygen [26], hydrogen peroxide [27], temperature [28], and pH. [29] In particular, wound pH has been widely reported as a promising infection biomarker, given the significant alkaline shift that takes place following infection onset. [30,31,32,33]

Many pH-monitoring devices use electrochemical sensors. Conventional glass-based sensors were used [34], however, these tend to be mechanically fragile and lack the flexibility required for clinical handling and dressing conformability to the wound. [35] To overcome these issues, other sensing electrode materials have subsequently been investigated, such as metal oxides [36], metal/metal oxides [37], and polymers/carbon [38,39], with limitations in the fabrication stage, leakage of the liquid electrolyte [40], prolonged activation time (~80 minutes), and low sensor stability. [39] Ion-sensitive field-effect transistors (ISFETs) [41,42] have also been applied with promising experimental data, although risks persist for the patients, including the high operating voltage, the output current drift, and the device stability in an aqueous environment. [43] Aiming at patient-friendly sensors, Laser-Induced Graphene (LIG) sensors have been produced that are flexible and can conform to the wound bed, although their fabrication is limited [44] and their pH sensing capability is yet to be fully realised. [45,46] Furthermore, bending the film also leads to changes in electrical resistance and negatively impacts the chemical sensing performance, which is a barrier to clinical applicability. [47] Another complication of electrochemical sensors of all types is they often rely on a bulky battery, which can negatively affect the wound environment. [48,49]

Alternative approaches to pH monitoring include colorimetric methods [50,51], fluorometric methods [52,53], and the use of optical fibre-based sensors. [54] The most common fibre optic systems for biomarker monitoring typically employ dye indicators or fluorophores deposited on tip of an optical fibre for pH sensing. [55] The main drawback of optical chemical sensors is the limited long-term stability, which can be caused by photobleaching or leaching of the dye. [56] Panzarasa et al. utilised pyranine, a non-toxic fluorescent dye, to detect changes in pH within the physiological pH window. [52] Limitations of the system include the need for extra filters to minimise dye leaching, limited pH range sensitivity, and the fact that a UV lamp is required to get pH readings. Although there are several fibre systems utilising halochromic dyes, challenges remain regarding how to ensure long-lasting dye retention, durable sensing functionality and minimal impact on the wound environment. Many systems display significant dye leaching [57,58], whereas systems utilising dye immobilisation in polymer matrices can cause alterations in the halochromic behaviour of the dyes, so the sensing capability at the target pH range may be affected. [59,60,61] Consequently, the translation of these smart dressings from the bench to the bedside is still in its infancy [10], and significant testing is required to ensure that the prolonged application of such devices does not elicit adverse effects on the wound healing process, skin cells, or lead to discomfort for users. [62]

This work aims to address the need to integrate an easy-to-apply therapeutic wound dressing with visual infection-sensing capability to reduce healthcare costs and limit the misuse of antibiotics. Unlike previous electrochemical sensors, which generate large quantities of data, which must be analysed, this dressing is designed to be simple enough for a patient to detect infection (by eye) at home by way of a safe, rapid colour change. Having previously demonstrated the wound healing capability of a collagen-based hydrogel *in vivo* [63,64,65], we hypothesised that a passive, visual early warning infection diagnostic system could be integrated through the application of a halochromic dye, i.e. bromothymol blue (BTB). Two different integration strategies, i.e. drop-casting and electrospinning, were pursued to ensure long-lasting dye retention in the dressing matrix, compatibility with the wound environment, and regulatory-friendly medical device classification. Experimental work was therefore carried out to assess dye loading, dye retention, and material colour change in simulated wound fluids. The microstructure of resulting prototypes was inspected by electron microscopy and complemented by analytical techniques and rheology to investigate the effect of BTB encapsulation on material properties. Ultimately, the impact of the dressing on cellular activity was analysed via culture of L929 murine fibroblasts with sample extracts.

## 2 Experimental

### 2.1 Materials

Rat tails were provided post-mortem by the University of Leeds (UK) and employed to extract type I collagen via acidic treatment. [64] Polysorbate 20 (PS-20), 4-vinylbenzyl chloride (4VBC), triethylamine (TEA), poly(ε-caprolactone) (PCL, $M_n$: 80,000 g·mol$^{-1}$) and N,N-Dimethylformamide (DMF) were purchased from Sigma-Aldrich (Gillingham, UK). 2-Hydroxy-4'-(2-hydroxyethoxy)-2-methylpropiophenone (I2959), and 1,1,1,3,3,3-Hexafluoro-2-propanol (HFIP) were purchased from Fluorochem Limited (Glossop, UK). Absolute ethanol (EtOH) was purchased from VWR international. A 6 N Hydrochloric acid solution was purchased from Thermo Fisher Scientific and diluted with deionised water before use.

Poly(methyl methacrylate-co-methacrylic acid) (PMMA-co-MAA) ($M_w$: 125,000 g·mol$^{-1}$; [MAA]:[MMA] = 1:2) was kindly donated by Evonik Rohm GmbH (Weiterstadt, Germany). Bromothymol Blue (BTB) was purchased from Alfa Aesar (Heysham, UK).

Dulbecco's modified eagle medium (low glucose, DMEM), trypsin, foetal bovine serum (FBS), and penicillin-streptomycin were purchased from Sigma. Phosphate buffered saline (PBS) was purchased from Lonza (Slough, UK). Calcein-AM/ethidium homodimer Live/Dead assay and alamarBlue™ Cell Viability Reagent were purchased from Thermo Fisher. Calcein-AM and ethidium homodimer was diluted to 2 and 4 μM, respectively, before use. All other chemicals were purchased from Sigma-Aldrich unless specified.

### 2.2 Preparation of drop-cast collagen-based materials

The collagen-based contact layer of the dressing prototype was realised as previously reported. [64] Briefly, type I Collagen extracted in-house from Rat Tails (CRT) was dissolved in a 10 mM hydrochloric acid solution (0.25 wt.%, 10 mM HCl) under magnetic stirring at room temperature and the pH neutralised to pH 7.4. Following addition of PS-20, 4VBC, and TEA, the reaction mixture was stirred for 24 hours at room temperature, prior to precipitation in a 10-volume excess of pure ethanol. The functionalised collagen product (4VBC) was recovered by centrifugation (10,000 rpm, 30 min, 4 °C) and air-dried.

Films were prepared by dissolving the 4VBC-reacted collagen product (0.8 wt.%) in I2959-supplemented solutions of 10 mM HCl (1 wt.% I2959). The solution was cast onto a 12-well plate (1.2 g of solution per well) and UV irradiated at 365 nm (Chromato-Vue C-71, Analytik Jena, Upland, CA, USA) for 15 min on both the top and bottom sides. UV-cured collagen hydrogels were thoroughly washed with deionised water and dehydrated in an increasing series of ethanol-deionised water solutions (0, 10, 20, 40, 60, 80, 3 × 100 vol.%). The resulting film was air-dried prior to further use. Cast films prior to and following UV curing were coded as TF and TF*, respectively. For the preparation of Freeze-Dried samples (FDs), the previously mentioned I2959 supplemented solution of functionalised collagen (0.8 wt.% 4VBC, 1 wt.% I2959) was poured into a 12-well plate (1.2 g of solution per well), covered in aluminium foil and freeze-dried using an Alpha 2-4 LDplus (Martin Christ Gefriertrocknungsanlagen GmbH, Osterode am Harz, Germany). UV-cured freeze-dried samples were coded as FD*.

After thorough vortexing, a drop of up to 100 μl of BTB solution (0.2 wt.% BTB in deionised water) was applied to the dry UV-cured samples and exposed to air for up to 12 hours. The resulting BTB drop-cast samples of TF* and FD* were coded as either D-TF* or D-FD*.

### 2.3 Preparation of two-layer electrospun constructs

Aforementioned samples of FD* were used as a fibre collector during electrospinning, to generate a two-layer composite structure of freeze-dried collagen and dye-encapsulated fibres. Electrospinning solutions were prepared in either HFIP (6 wt.% PCL) or a mixture of EtOH and DMF (2:1 ratio, 15 wt.% PMMA-co-MAA). [32] BTB was added (0.5 wt.% of the polymer weight) while stirring the solution magnetically.

BTB-loaded solutions of PCL were electrospun with an applied voltage of 16 kV, a flow rate of 1.8 ml·h$^{-1}$, and a working distance of 10 cm. BTB-loaded solutions of PMMA-co-MAA were electrospun with an applied voltage of 11 kV, a flow rate of 0.5 ml·h$^{-1}$, and a working distance of 18 cm. Constructs based on BTB-encapsulated fibres of either PMMA-co-MAA or PCL were coded as C-PMMA-co-MAA and C-PCL, respectively. The BTB-encapsulated fibres of either PMMA-co-MAA or PCL were coded as D-PMMA-co-MAA and D-PCL. The BTB-free fibre controls were coded as either F-PMMA-co-MAA or F-PCL.

### 2.4 Circular dichroism

Circular dichroism (CD) spectra were acquired using a Chirascan CD spectrometer (Applied Photophysics Ltd., Leatherhead, UK). The I2959-supplemented solution of functionalised collagen (0.8 wt.% 4VBC, 1 wt.% I2959) was loaded into a Type 106 100 μm demountable cell (Hellma, Müllheim, Germany), whereby BTB (30 μl, 0.2 wt.% in deionised water) was then added. The sample was cured under UV for 30 minutes and loaded into the Chirascan CD spectrometer. CD spectra were obtained with a 3 nm bandwidth and a 20 nm·min$^{-1}$ scanning speed. A BTB-free sample was prepared and analysed as a control.

### 2.5 Differential scanning calorimetry

Differential scanning calorimetry (DSC) was conducted on a DSC Q100 (TA Instruments, Newcastle, DE, USA), temperature scans were conducted in the range of -10-90 °C with a 10 °C·min$^{-1}$ heating rate on both drop-cast samples and BTB-free controls. The DSC cell was calibrated using indium with 20 °C·min$^{-1}$ heating rate under 50 cm$^3$·min$^{-1}$ nitrogen atmosphere. 5-10 mg of sample were applied in each measurement.

### 2.6 Attenuated total reflection Fourier transform infrared (ATR–FTIR) spectroscopy

ATR–FTIR spectra were recorded on dry samples using a Spectrum One FT-IR Spectrometer (PerkinElmer, Waltham, MA, USA) with a Golden Gate ATR attachment (Specac Ltd., London, UK). Scans were conducted from 4000 to 600 cm$^{-1}$ with 100 repetitions averaged for each spectrum. Resolution and scanning intervals were 4 cm$^{-1}$ and 2 cm$^{-1}$, respectively.

### 2.7 Rheology

Rheological data from both BTB-free and drop-cast samples of TFs were recorded following rehydration in deionised water using an MCR 301 rheometer (Anton Paar, Graz, Austria). An amplitude sweep was initially conducted to identify the linear-elastic region using an angular frequency of 10 rad·s$^{-1}$, then a frequency sweep was conducted on a fresh sample at room temperature with a constant amplitude of 1% and angular frequencies from 100 to 0.1 rad·s$^{-1}$.

### 2.8 Quantification of BTB loading

To quantify the BTB loading content of drop-cast samples, a gravimetric method was employed (n=9). The mass of the individual samples ($m_i$) was recorded using a precision balance before the BTB solution (100 μl, 0.2 wt.% BTB) was cast using a micropipette. After air-drying the samples for 48 hours, the final mass ($m_f$) was recorded using a precision balance and the loading efficiency (LE) was calculated using Equation 1:

$$LE = \frac{(m_f - m_i)}{m_{BTB}} \times 100 \qquad (1)$$

where $m_{BTB}$ is the mass of BTB contained in the volume of the aqueous solution applied to the samples.

To quantify the BTB loading in electrospun fibres (n=9), UV-vis spectroscopy was employed using a UV-Vis spectrophotometer (Model 6305, Jenway, Dunmow, UK). A calibration curve was built for each fibre group by recording the absorbance at 432 nm; either ethanol or acetone was selected as the solvent of choice for BTB-containing fibres of PMMA-co-MAA and PCL, respectively. The calibration curves and equations are shown in **Figure S1**.

**2.9 Dye release measurements**

Dye retention was indirectly assessed by measuring the release of the dye out of the dressing prototypes. Individual samples (n=3) containing up to 220 µg of BTB were incubated at room temperature in 5 mL of McIlvaine solution adjusted to either pH 5 or pH 8. Samples labelled as 'Sub' were submerged in 6-well plates (Corning, Corning, NY, USA) containing the McIlvaine solution, whereas the remaining samples were placed in Petri dishes (Corning, Corning, NY, USA) onto a layer of McIlvaine solution, simulating a dressing placed on a moist wound. Over a four-day period, the amount of BTB in each solution was periodically determined via UV-Vis spectrophotometry of the solution. Calibration curves were produced using two McIlvaine solutions, at pH 5 and pH 8, with recordings taken at 432 nm and 616 nm, respectively. Resulting absorbance data were used to derive the amount of BTB released from the samples at each time point and solution pH. The calibration curves and equations are shown in **Figure S2**.

**2.10 Colorimetry**

The colour of the dressings was recorded using an SF600 Plus-CT spectrophotometer (Datacolor, Lucerne, Switzerland). After calibration to standard black/white standards, samples of TF*, FD*, C-PCL, and C-PMMA-co-MAA were loaded into the Datacolor to be analysed. The Lightness, Chroma, and Hue were all recorded at 7 different locations using a 3 mm window on each sample to get an average reading.

**2.11 Scanning electron microscopy**

Samples were inspected via scanning electron microscopy using a Hitachi S-3400N (Hitachi, Tokyo, Japan). All samples were gold-sputtered using an Agar Auto sputter coater (Agar Scientific, Stansted, UK) prior to examination. The SEM was fitted with a tungsten electron source and the secondary electron detector was used. The instrument was operated with an accelerating voltage of 3 kV in a high vacuum with a nominal working distance of 10 mm. The cross-sectional morphology of BTB-encapsulated constructs was investigated after freeze-fracturing in liquid nitrogen and lyophilisation. Electrospun fibre and hydrogel pore diameters were recorded manually using ImageJ software, with 50 repeats.

**2.12 Extract cytotoxicity study**

To evaluate the cytotoxicity of the BTB-encapsulated samples, 48-hour extracts from either thin films or freeze-dried samples were tested with L929 murine fibroblasts and a resazurin assay and live/dead assay were utilised. The L929 fibroblasts were seeded in a 24-well plate with a seeding density of $1 \times 10^4$ cells per ml, with 1 ml in each well. Test wells contained extracts from BTB-loaded samples. Extracts of samples of FD* and TF* were obtained following incubation in McIlvaine solutions of either pH 5 or pH 8 (48 hours, 37 °C). BTB was then added to 1 ml of the resulting extract to produce solutions with a BTB concentration of 3.2 mM. 40 µL of this extract solution was added to each of the test wells, which contained an excess of BTB with respect to the values seen from the dye release measurements.

For the resazurin assay, after incubation for 48 hours, the media was aspirated off, wells were washed with PBS, and fresh media was added with 10 vol% alamarBlue solution. After briefly shaking, the well plates were incubated for a further 6 hours. Cell viability was determined by incubating 150 µl of each well solution in a 96-well plate for fluorescence analysis, with an excitation wavelength of 560 nm and an emission wavelength of 590 nm, using Equation 2:

$$\% \ cell \ viability = \frac{Fl_{590} \ Extract}{Fl_{590} \ Control} \times 100 \qquad (2)$$

For the live/dead stain, after incubation for 48 hours, the medium was aspirated off, wells were washed twice with PBS, and 100 µL of premixed calcein-AM and ethidium homodimer stain were added. Well plates were then incubated for a further 45 minutes before being washed twice in PBS and imaged under a confocal microscope using a TCS SP8 (Leica, Wetzlar, Germany).

**2.13 Statistical analysis**

Data normality tests were carried out using OriginPro 8.51 software (OriginPro, OriginLab Corporation, Northampton, MA, USA). Statistical differences were determined by one-way ANOVA and the post hoc Tukey test. A *p* value lower than 0.05 was considered significantly different. Data are presented as mean ± SD.

**3 Results and discussion**

In the following, the design, manufacture and characterisation of a theranostic wound dressing will be presented, aiming to safely integrate an infection-sensing colour change functionality with the previously demonstrated wound healing capability of a photoinduced collagen film (**Figure S3**). [63]

**3.1 Design and microstructure of theranostic dressings**

**Figure 1A** shows how the structure and colour of the halochromic dye bromothymol blue (BTB) change with pH. In an acidic environment (pH <7), the dye is yellow in colour and can reversibly shift to a blue colour if exposed to an alkaline medium (pH >7). This colour shift is associated with a variation in molecular configuration, whereby, below pH 7, BTB presents a monovalent anion with the sulfonate group; above pH 7, proton dissociation from the phenolic group results in a bivalent anion and increased negative electrostatic charge. [66]

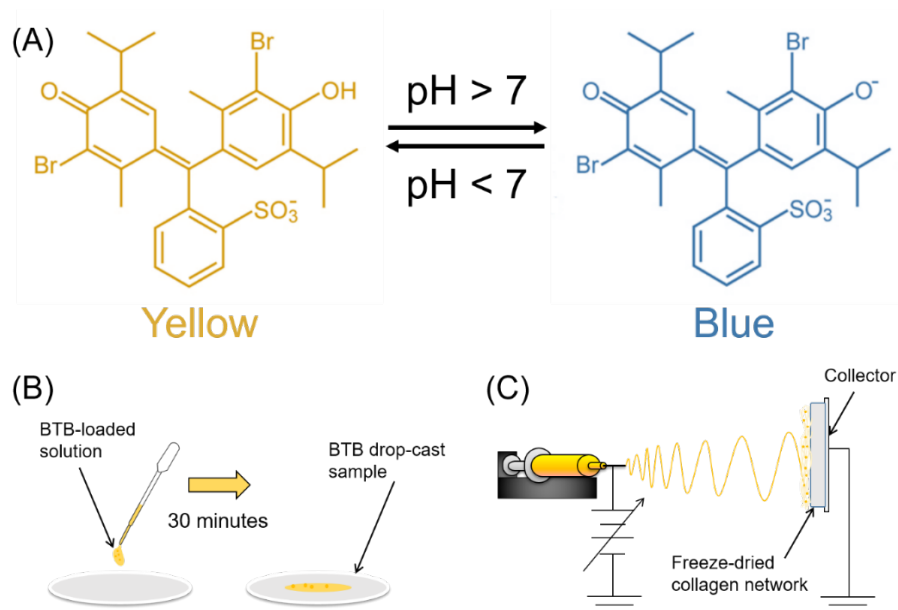

**Figure 1.** Research strategies pursued for the realisation of a theranostic wound dressing. (A): The reversible colour change capability of BTB is leveraged for the introduction of visual infection-sensing capability in the collagen-based dressing; (B): Schematic of the localised BTB drop-casting route yielding the therapeutic and infection-sensing dressing; (C): Schematic of the electrospinning process yielding BTB-loaded fibres onto the freeze-dried collagen network. The resulting dual-layer composite dressing consists of a therapeutic collagen-based wound contact layer and an infection-sensing top fibrous layer.

Leveraging this reversible molecular rearrangement, two manufacturing routes, i.e. drop-casting (Figure 1B) and electrospinning (Figure 1C), were pursued to accomplish a theranostic dressing enabling infection-sensing colour change and ensure retention of the dye in the wound dressing structure. The drop-casting method aims to produce a localised concentration of BTB onto the dry collagen film, whereby a small volume of a BTB-loaded aqueous solution is delivered onto the sample, followed by solvent evaporation for at least 30 minutes. Here, the negatively charged configurations of BTB are expected to generate electrostatic complexation with the remaining positively charged primary amino terminations of the collagen matrix [64], aiming to accomplish dye confinement in the dressing structure. On the other hand, electrospinning is employed to generate a two-layer composite, whereby the freeze-dried collagen network serves as the fibre collector, aiming to create a homogeneous layer of BTB-encapsulated fibres. With this strategy, the underlying collagen layer is intended to act as both the wound dressing contact layer, aiming to support wound healing *in situ*; as well as a structural barrier, to ensure minimal diffusion of the dye away from the dressing towards the wound. Overall, the two manufacturing routes are therefore expected to provide the resulting dressing with an integrated colour change capability through contact with the infected wound exudate (i.e., at pH >7), whereby either the molecular scale (electrostatic interactions) or the microscale (fibrous layer) are leveraged to ensure long-lasting dye retention.

Following the manufacture of the theranostic prototypes, the microstructure of both the dressing film and electrospun construct was investigated (**Figure 2**). The cross-sectional morphology of a previously freeze-fractured sample was initially visualised using SEM to see if the addition of BTB influenced the matrix morphology. A series of irregular pores were observed (Figure 2A-B) typical of a crosslinked collagen network [67], with a pore diameter of 33±12 μm and 32±10 μm (n=50) for the BTB-free controls (TF*) and the drop-cast films (D-TF*), respectively. There was no statistical difference in pore diameter between the two samples, therefore it was concluded that the addition of BTB does not affect the microstructure of the film dressing.

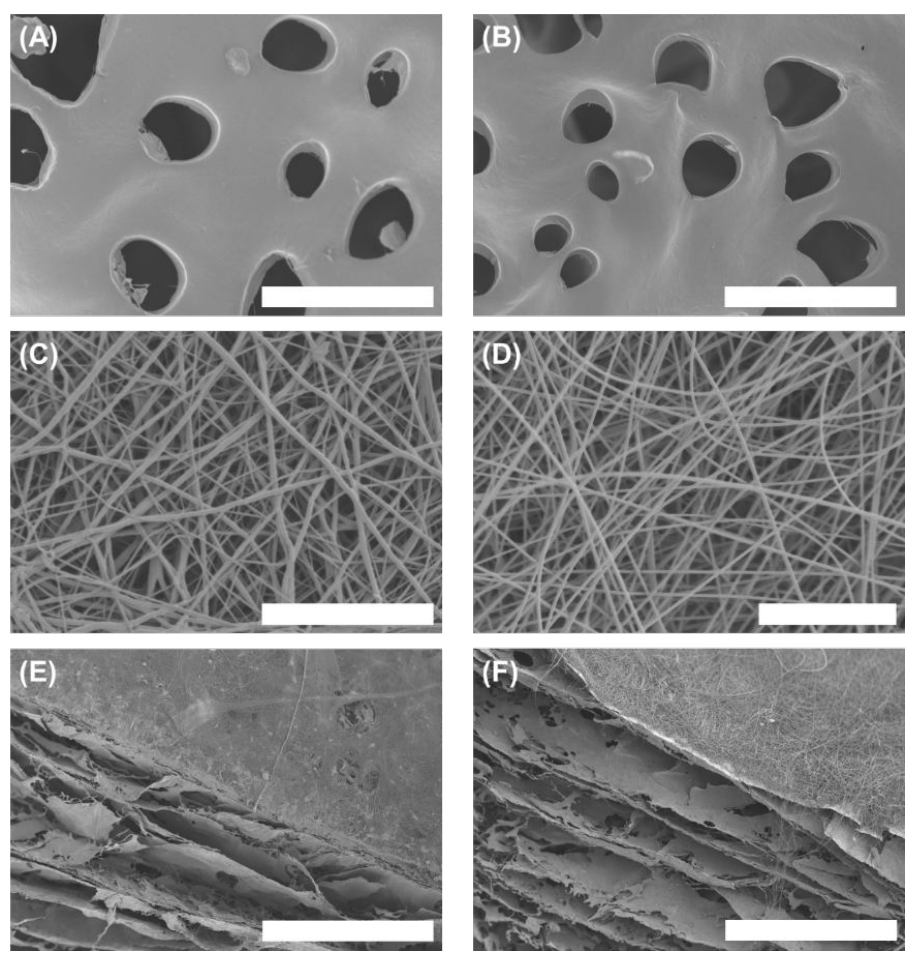

**Figure 2.** SEM micrographs of wound theranostic prototypes. (A-B): freeze-fractured samples of TF* (A) and D-TF* (B). Scale bar: 100 μm. (C-D): fibres of D-PCL (C) and D-PMMA-co-MAA (D). Scale bar: 20 μm. (E-F): electrospun construct of C-PCL (E) and C-PMMA-co-MAA (F). Scale bar: 500 μm.

Together with the drop-cast samples, the BTB-encapsulated electrospun fibres were analysed to check the impact of BTB on fibre morphology and size. BTB-loaded solutions of PMMA-co-MAA and PCL generated uniform electrospun fibres with minimal bead formation, with comparable fibre diameters to those seen in previously electrospun PCL meshes. [68,69] The average diameter of D-PCL fibres was 507±166 nm, which proved to be comparable to the one measured in D-PMMA-co-MAA fibre (Ø =596±129 nm, n=50), in line with the minimal effect of soluble factor loading on fibre diameter. [70,71,72] Other than the individual mesh, electrospinning of the above-mentioned fibres onto the collagen film was successfully demonstrated by SEM, whereby the two-layer dressing structure consisting of the freeze-dried collagen wound contact layer below a dense mat of nonwoven fibres is clearly visible on both constructs of C-PCL and C-PMMA-co-MAA (Figures 2D-E).

**3.2 Loading efficiency, infection-sensing colour change and dye retention capability**

Following microstructure investigations, the BTB loading efficiency in the samples generated via both manufacturing strategies was studied via either UV-Vis spectroscopy or gravimetric analysis. Both samples revealed an average loading efficiency of 99 wt.% (**Table 1**), suggesting that both drop-casting and electrospinning are viable methods for integrating BTB into the theranostic wound dressing.

**Table 1.** Quantification of BTB loading in drop-cast films and electrospun constructs. *LE*: loading efficiency; *L×C×h*: Luminance (L), Chroma (C) and Hue (h) recorded following sample incubation in both acidic (pH 5) and alkaline (pH 8) media.

| Sample ID | LE /wt.% | L×C×h pH 5 | L×C×h pH 8 |
|---|---|---|---|
| D-TF* | 99±5 | 60×60×72 | 30×14×141 |
| D-FD* | 99±3 | 73×65×71 | 46×39×82 |
| C-PCL | 99±4 | 91×48×86 | 59×52×93 |
| C-PMMA-co-MAA | 99±1 | 90×42×86 | 59×43×94 |

Consistent with the high loading efficiency, all samples displayed a colour change following incubation at pH 5 and pH 8 (**Figure 3A-B**), which was visible by the eye even after 1 minute of response time (**Figure S4**). This colour change proved to be reversible over multiple cycles, so that the yellow samples characteristic of an acidic environment switched to a blue colour when transferred to an alkaline solution (Figure 3C). Quantitatively, the LCh colour values shown in Table 1 demonstrated a colour shift when the samples were moved from an acidic to an alkaline environment, consistent with a decrease in Lightness and a shift in Chroma, away from values associated with yellow and towards values representing blue.

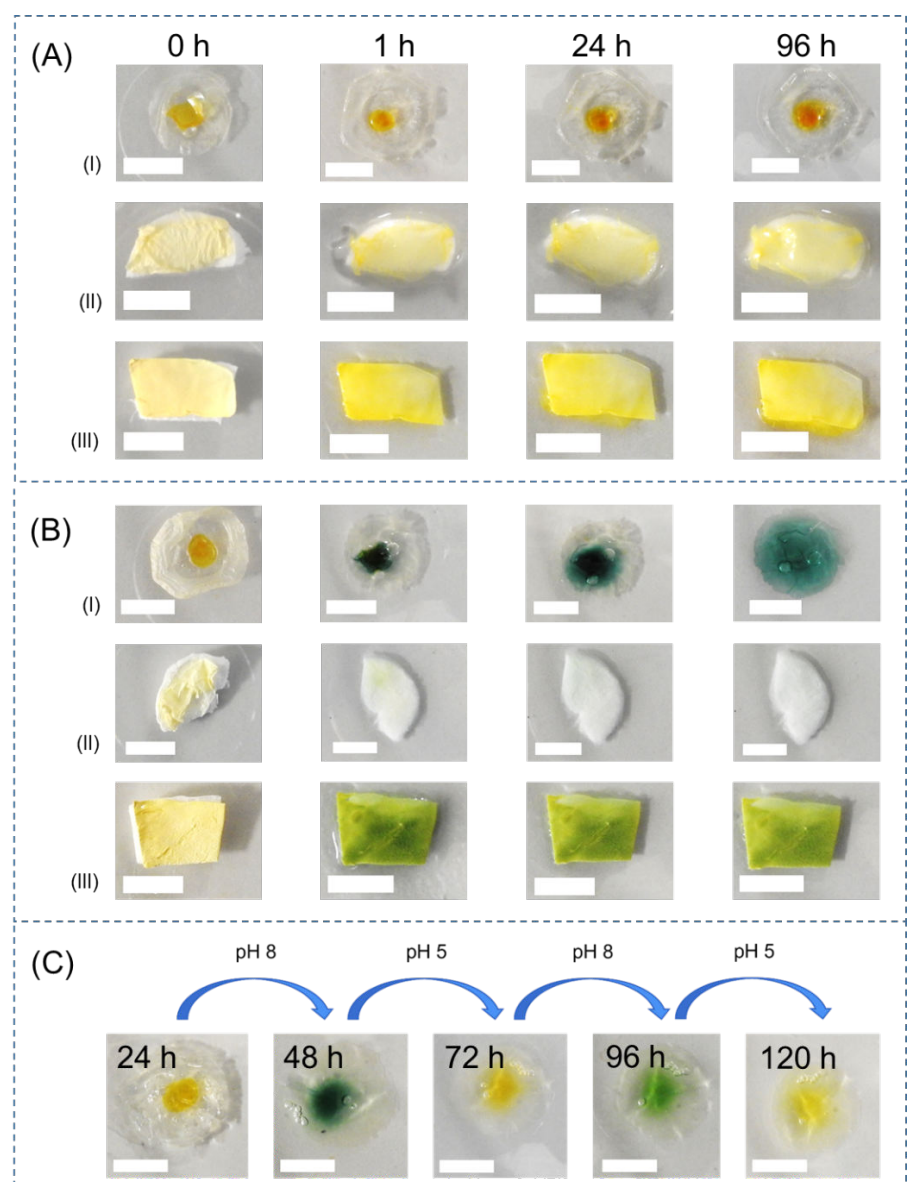

**Figure 3.** pH-induced colour change capability of BTB-loaded materials. (A-B): Images of drop-cast *film* (I), BTB-loaded electrospun PMMA-co-MAA *construct* (II), and BTB-loaded electrospun PCL *construct* (III) measured over 96 hours in McIlvaine buffer at pH 5 (A) and pH 8 (B). (C): Reversible colour shift and durable infection sensing capability of drop-cast films over 120 hours following alternating incubation steps in pH 5 and pH 8 McIlvaine solutions. Scale bars 1 cm.

Having confirmed the dye loading efficiency and dressing colour change, it was critical to confirm that the dye was retained in the structure over a 96-hour incubation in simulated wound fluids, where 96 hours was selected as a clinically relevant dressing application time [73]. Retention of the dye in the dressing is key to ensure long-lasting infection-sensing capability and minimal risks of dressing-induced alteration of the wound environment. The drop-cast samples revealed significantly lower release values with respect to the dual layer constructs (**Figure 4A-B**), indicating that BTB was retained by the collagen matrix, an observation that agrees with the previously-observed localised distribution of the dye in the former samples (Figure 3).

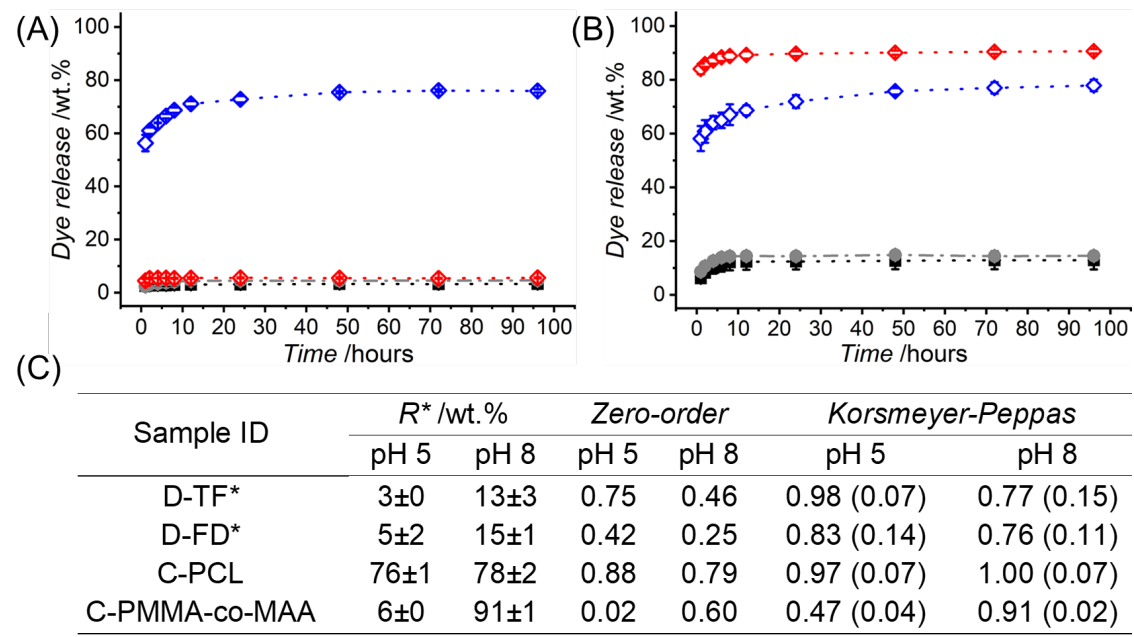

| Sample ID | $R^*$ /wt.% | | Zero-order | | Korsmeyer-Peppas | |
|---|---|---|---|---|---|---|
| | pH 5 | pH 8 | pH 5 | pH 8 | pH 5 | pH 8 |
| D-TF* | 3±0 | 13±3 | 0.75 | 0.46 | 0.98 (0.07) | 0.77 (0.15) |
| D-FD* | 5±2 | 15±1 | 0.42 | 0.25 | 0.83 (0.14) | 0.76 (0.11) |
| C-PCL | 76±1 | 78±2 | 0.88 | 0.79 | 0.97 (0.07) | 1.00 (0.07) |
| C-PMMA-co-MAA | 6±0 | 91±1 | 0.02 | 0.60 | 0.47 (0.04) | 0.91 (0.02) |

**Figure 4.** Dye release profiles recorded following room temperature incubation of theranostic prototypes in a McIlvaine buffer at pH 5 (A) and pH 8 (B). (··■··): D-TF*; (─●─): D-FD*; (··◇··): C-PCL; (··◇··): C-PMMA-co-MAA. Quantification of BTB release in drop-cast films and electrospun constructs. (C): $R^*$: BTB fraction released following 96-hour-incubation at pH 5 and pH 8. Zero-order: coefficient of determination ($R^2$) values obtained via zero-order release data fitting; Korsmeyer-Peppas: $R^2$ values following release data fitting via Korsmeyer-Peppas model (number in bracket indicates the value of the release exponent).

As presented in Figure 4C, samples of D-TF* indicated a 96-hour averaged release of 3 wt.% and 13 wt.% in acidic and alkaline environments, respectively, with similar trends observed with the freeze-dried samples D-FD* (pH 5: $R^*$= 5±2 wt.%; pH 8: $R^*$=15±1 wt.%). Given that the aforementioned samples were drop-cast with different dye quantities (i.e. either with 200 µg for testing at pH 5 or 60 µg for testing at pH 8), a release of up to 10 µg dye was observed regardless of the pH tested. The reason for the aforementioned discrepancy in BTB loading for samples tested at pH 5 and pH 8 is due to the different calibration curves and UV-vis spectroscopy detection limits (Figure S2). The significant retention of the dye by the dressing is attributed to the development of electrostatic interactions between the BTB and the primary amines in the collagen-based hydrogel. The protonated amine groups may also react with the sulfonate group, a well-known leaving group [74,75], on the BTB to form covalent bonds, with the release of $H_2SO_3$, [76] though this is not supported by the release profiles presented in Figure 4.

As shown in Figure 1A, BTB bears either one or two negatively charged groups, which can interact with any remaining protonated amines ($pK_a$ ~9) in the collagen hydrogel ([Lys] ~1.6×10$^{-4}$ mol·g$^{-1}$). [64] Evidence that the dye retention mechanism is through these amine-dye interactions is seen in Figure 3A-B. The dye is found to spread throughout the hydrogel at pH 8, where fewer protonated amines are expected in the collagen material. In contrast, the same dressing exhibits a localised dye distribution at pH 5, in light of the increased content of protonated amines. The significant retention of the dye by the collagen film is further supported by the reversible colour shift observed following alternating sample incubation steps in new acidic and alkaline solutions over a 120-hour time period (Figure 3C). These observations highlight the durability of the infection-sensing capability and minimise the risk that a solubility limit was reached by the dye in the aqueous solution.

**3.2.1 Dye release testing in acidic conditions**

Under acidic simulated healthy wound conditions (pH 5), samples of C-PCL proved to release over 50 wt.% of the loaded BTB after the first hour (Figure 4A), compared to just 4.5 wt.% and 2.5 wt.% of dye released from samples C-PMMA-co-MAA and drop-cast samples, respectively. After the 96-hour incubation period, 76 wt.% of BTB was released from samples C-PCL, compared to just 3 wt.% from TF*, 5 wt.% from FD* and 6 wt.% from C-PMMA-co-MAA. The drop-cast samples D-TF* and D-FD*, therefore, enable a 90 % reduction in dye release compared to PCL-based samples, an observation that supports the previously described electrostatic complexation mechanism between BTB and the collagen hydrogel. Accordingly, these results speak against the fact that the observed dye retention in the drop-cast collagen samples is due to a solubility limit being reached by the released dye in the McIlvaine solutions.

The aforementioned dye release values measured with the drop-cast samples compare favourably to a number of previous systems seen in the literature. A natural anthocyanin halochromic dye has previously been incorporated into gellan gum-based composite films, however, the composite films liberated the anthocyanins into media at pH 2, 6, and 7.4. [77] Previous attempts to incorporate pH-sensitive dyes using electrospinning have also led to extensive dye leaching. Nitrazine Yellow was electrospun with polyamides to form nanofibres, and after 24 hours dye release was observed at all pH values, peaking at 70-87 % at pH 12. [78] After the addition of a complexing agent, there was still up to 40 % dye release under the same conditions. [78] Nitrazine Yellow has also been electrospun into PCL and PCL/chitosan fibres with varying release values after 24 hours at pH 8, up to 57 %. [57] In this case, the addition of a complexing agent reduced the dye release values to 0-7 %, however, the complexing agent caused alterations in the halochromic behaviour of the dye. [57]

The fast release results from D-PCL samples agree with previous work investigating the release of Rhodamine B from PCL and PCL-based electrospun fibres, whereby a large initial burst release was observed. [79] At both room temperature and physiologically relevant temperatures, PCL chains are in a rubbery state ($T_g$ ~-60 ºC) [80], meaning that the polymer chains have increased mobility so that diffusional release of additives is promoted. This observation is also supported by the close fitting of the release data of C-PCL with the Korsmeyer-Peppas model ($R^2$ =0.97, Figure 4C), whereby a release exponent lower than 0.5 was determined in both experimental conditions, in line with a diffusion-driven release mechanism. Unlike PCL, the low release of BTB observed in the acidic medium with samples of C-PMMA-co-MAA is mainly attributed to the incompatibility of PMMA-co-MAA with water at pH <7 [81], as well as its high glass transition temperature ($T_g$ ~170°C). [82] The polymer chains of PMMA-co-MAA are therefore expected to be more rigid and immobile compared to the ones of PCL in the tested experimental conditions, leading to a slower rate of diffusion from the fibres [82], as indicated by the low fitting ($R^2$ ≥0.45) of the release data with the Korsmeyer-Peppas model (Figure 4C).

No statistical difference in dye release was observed after 96 hours (pH 5) between drop-cast samples that were left to dry for 30 minutes and those left for 12 hours ($p$ > 0.05) (**Table S1** and **Figure S5**). For this reason, further analysis was solely conducted on samples produced using the less time-intensive method. There was also no statistical difference in dye release after 96 hours between the thin film and freeze-dried variants ($p$ > 0.05), further indicating that the reduction in dye release is a function of molecular interactions rather than an aspect of the sample geometry. Samples of D-TF*, D-TF*12, D-FD*, and D-FD*12 all had mean dye release values after 96 hours that were statistically insignificant ($p$ > 0.05) when compared to D-PMMA-co-MAA (Figure S5), which is known to provide a low rate of diffusion due to its glass transition temperature and rigid chains.

It was also of interest to determine the extent of dye release when samples were completely submerged in the simulated wound fluid (pH 5), compared to previously discussed release experiments where samples were in contact with, but not submerged by, the medium, as a more clinically descriptive scenario of the dressing environment. As expected, significant differences were observed with samples D-FD* and D-TF*, on the one hand, and their submerged counterparts, on the other hand (**Tables S2 and S3**), although only up to 30 µg of BTB were released by the submerged samples after 96 hours, equating to a 15 wt.% of dye release. Given that this amount of BTB is soluble in water in selected volumes (5 mL), this relatively low value of dye release further supports the previous explanation on the development of molecular interactions between the protonated amines of the photoinduced collagen network and the negatively charged dye molecules, which keep the dye incorporated with the collagen hydrogel.

In terms of the fibrous constructs, the C-PCL samples afforded a statistically significant reduction in dye release after 96 hours when compared to the D-PCL fibres ($p > 0.05$, Figure S5, Table S1), indicating that the collagen wound contact layer retards the release of BTB from the top fibrous layer. On the other hand, there was no significant difference in dye release between C-PMMA-co-MAA samples and the D-PMMA-co-MAA fibres after 96 hours ($p > 0.05$), which agrees with the incompatibility of this polymer in acidic aqueous environments.

### 3.2.2 Dye release testing in alkaline conditions

Under alkaline simulated infected wound conditions, sample C-PMMA-co-MAA had significant dye release (84 wt.% after 1 hour), in line with the well-known solubility switch of PMMA-co-MAA above pH 7 [83], and the consequent complete solubilisation of the corresponding fibres. As was the case under acidic simulated healthy wound conditions (pH 5), samples of C-PCL released over 50 % of the loaded BTB after just 1 hour. Here, the movement of the rubbery PCL chains enables the dye to diffuse out of the fibres and into the McIlvaine solution, as supported by the close fitting of release data by the Korsmeyer-Peppas model and low release exponent ($R^2 = 1$, $n = 0.07$; Figure 4C). In comparison, the dye release is significantly lower in the drop-cast samples, with only up to 9 wt.% dye release observed after 1 hour (Figure 4C). After the 96-hour incubation period, 76 wt.% and 91 wt.% of BTB is released from C-PCL and C-PMMA-co-MAA, respectively, in contrast to the 13 wt.% and 15 wt.% release observed with samples of TF* and FD*, respectively. Consequently, over 80 % reduction in dye release is afforded by the drop-cast samples compared to both C-PCL and C-PMMA-co-MAA samples.

As was the case with samples examined under acidic simulated healthy wound conditions, no statistical difference in dye release was observed after 96 hours in the near-infected alkaline environment between drop-cast samples that were left to dry for 30 minutes and those left for 12 hours ($p > 0.05$, Table S1). Drop-cast samples that were left to dry for 12 hours had mean dye release values after 96 hours that were statistically insignificant ($p > 0.05$) when compared to their submerged counterparts. In contrast, there was a statistically significant difference in dye release after 96 hours between drop-cast samples left for 30 minutes when compared to their submerged counterparts. Despite there being an increase in dye release when the samples were completely submerged, the maximum amount of dye released after 96 hours was ~14 µg (i.e. 24 wt.%), therefore the majority of the dye remained incorporated within the collagen network.

To further understand the release kinetics, the release data were fitted with the zero-order and Korsmeyer-Peppas models (Figure 4C). According to zero-order kinetics, a drug is delivered from the carrier at a constant rate, while a non-linear release is described by the Korsmeyer-Peppas model. The zero-order model had a good correlation ($R^2 > 0.7$) in the case of release from samples of TF* at pH 5 and from samples of C-PCL at both pH 5 and pH 8. Comparatively, the Korsmeyer-Peppas model had a good correlation for seven out of eight release data collected, with an average $R^2$ value of 0.84, and an $R^2$ value of 0.89 when the anomalous result is discounted (Figure 4C). The values of the release exponent, $n$, are also lower than one in all cases, indicating a reduced concentration gradient over time, thereby suggesting that the release follows first-order (or pseudo-first-order) kinetics, according to a diffusion-driven release.

### 3.3 Chemical composition and thermomechanical properties

ATR-FTIR spectroscopy was subsequently employed to elucidate the chemical composition of the drop-cast and electrospun samples and the development of any secondary interactions between the halochromic dye and the carrier polymers (**Figure 5**). The ATR-FTIR spectra of TF* and D-TF* confirmed the main spectroscopic bands associated with type I collagen (Figure 5A), i.e. amide I (1629-1655 cm$^{-1}$), amide II (1541-1598 cm$^{-1}$) and primary amines (3290-3320 cm$^{-1}$). Comparable ATR-FTIR spectra were also observed with samples of F-PCL, D-PCL, and C-PCL (Figure 5B), whereby the presence of the ester bonds was detected at 1719 cm$^{-1}$. At the same time, a slight red shift was detected in D-TF* samples compared to TF* samples. These red shifts occurred in the amide I peak, from 1655 to 1629 cm$^{-1}$, amide II peak, from 1598 to 1541 cm$^{-1}$, and primary amine peak, from 3320 to 3290 cm$^{-1}$. This observation supports the formation of secondary interactions, mainly hydrogen bonding and electrostatic interactions, between the collagen network and the BTB, in agreement with the retention of the dye in the network during the 96-hour time period investigated (Figure 3, Figure 4C, and Figure S5). This contrasts with the spectra seen in Figure 5B, where there is a lack of red shift after the encapsulation of BTB in the PCL fibres. This indicates the absence of secondary interactions, which is in agreement with the significant dye release revealed by the electrospun constructs (Figure 4).

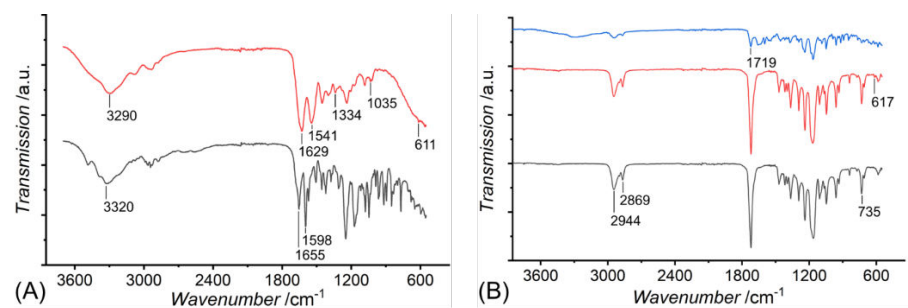

**Figure 5.** ATR-FTIR spectra of drop-cast samples (A) and electrospun construct (B). (A): TF* (black) and D-TF* (red). (B): F-PCL (black), D-PCL (red), C-PCL (blue).

The significantly increased dye retention in D-TF* samples in both acidic and alkaline simulated wound environments supported the further characterisation of this material aiming to establish the structure-property relations responsible for the above finding and assess its feasibility as a safe wound theranostic (**Figure 6**). Circular dichroism (CD) spectra were recorded for both TF* and D-TF* samples to elucidate the short-range protein organisation in the crosslinked network and discern any interactions between the dye and the protein (Figure 6A). Typically, CD spectra of type I collagen solutions display a positive peak at about 221 nm, which is attributed to the presence of right-handed triple helices, and a negative peak at about 197 nm, which is attributed to the left-handed polyproline-II chains. [84] Both these peaks were detected in the CD spectra of both TF* and D-TF* (Figure 6A), supporting the validity of CD in assessing the protein organisation in insoluble crosslinked networks [85,86], similar to the case of diluted collagen solutions. [64,87] Furthermore, the above results also suggest that the addition of BTB does not affect the 4VBC-functionalised triple helix organisation in the UV-cured collagen network. To quantify the triple helix organisation, the magnitude ratio between positive and negative CD peak intensities (*RPN*) of the TF* sample was measured to be ~12. Following dye drop-casting, the *RPN* was measured to be ~9, suggesting that there has been a 25% loss in collagen triple helix organisation, following UV curing in the presence of BTB. In principle, this could be attributed to the presence of electrostatic charges in the BTB molecule, potentially interfering with the secondary interactions mediating the folding of the collagen triple helix. On the other hand, Brazdaru et al. have previously shown that the *RPN* values can vary depending on the collagen concentrations [88], so any local variations in collagen density in the UV-cured crosslinked network can contribute to the aforementioned decrease in *RPN*.

Differential scanning calorimetry (DSC) was also conducted on TF* and D-TF* samples for further confirmation of dye-protein interactions. Figure 6B shows a thermogram of the two samples from -10°C to 90°C, whereby the lack of a large endothermic peak around 0°C suggests that the hydrogel samples were prepared correctly, i.e. as a water-swollen network with minimal excess of free water.

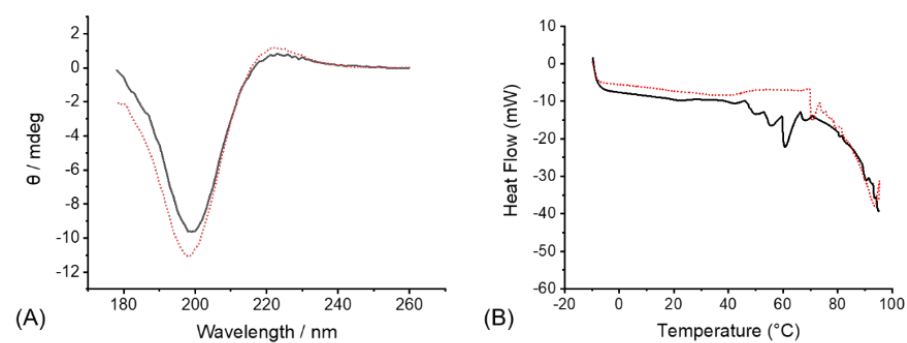

**Figure 6.** Far-UV CD spectra of ellipticity (*θ*) recorded at constant temperature (20 °C) (A), and DSC thermograms recorded at -10°C to 90°C (B) for TF* samples (black-solid) and D-TF* samples (red-dot).

Since the bound water forms hydrogen bonds with the collagen network rather than with other water molecules, the bound water is unable to freeze when the temperature is lowered below 0°C. [89] Other than that, a denaturation peak was seen in the TF* sample around 60°C, whereas the D-TF* sample presented a denaturation peak at 70°C, suggesting that the addition of the dye contributes an increased thermal stability of the protein. The long trailing endothermic peak from 75°C seen in the samples D-TF* is likely due to the onset of disassociation of the water molecules. Here, the presence of multiple hydrogen bonds in the hydrogel network delays the evaporation of water and is likely responsible for the broad endothermic transition over 100°C. [90]

Other than the thermal behaviour, the mechanical properties of TF* and D-TF* hydrogels were determined by rheology (**Figure 7**). The oscillatory shear data show that the storage modulus ($G'$) is dominant in both samples across the range of strain (Figure 7A) and frequencies (Figure 7B) observed, with values of $G'$ and loss moduli ($G''$) of up to ~8.5 kPa and ~1 kPa, respectively, in samples of D-TF*.

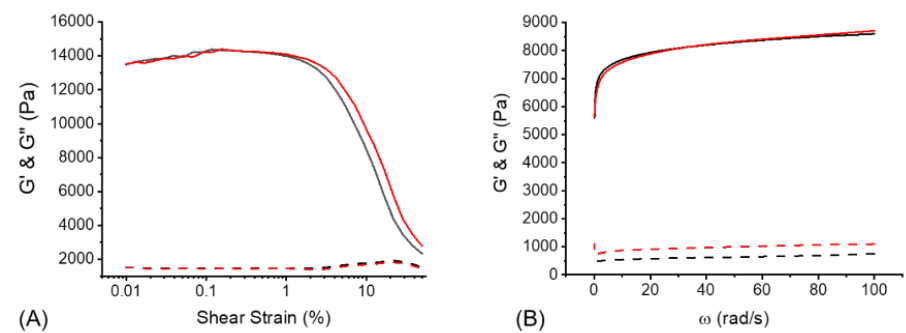

**Figure 7.** Amplitude (A) and frequency sweeps (B) of representative samples of TF* (black) and D-TF* (red). G': storage modulus (solid line); G": loss modulus (dashed line).

This more than 8-times increase in $G'$ compared to $G''$ indicates that both hydrogels behave as an elastomeric material regardless of the presence of BTB. This, together with the absence of a crossover frequency, agrees with the fact that both drop-cast and BTB-free hydrogels are permanently chemically crosslinked following UV curing. [81,87] The similarity of frequency sweep profiles in both samples of TF* and D-TF* reveals that the addition of BTB does not impact the rheological properties of the hydrogel. While the introduced BTB molecules have been shown to mediate dye-matrix secondary interactions (Figure 5A) leading to increased thermal stability (Figure 6B), these appear to be relatively weak or localised to impart a detectable change in macroscale mechanical properties.

### 3.4 Extract cytotoxicity study

A resazurin assay was subsequently conducted to evaluate the cytotoxicity of the drop-cast collagen samples. For completeness, the cytotoxicity of both D-TF* and D-FD* with murine L929 fibroblasts was examined over a 7-day cell culture period to determine if the processing route had any impact on cell viability (**Figure 8**). Extracts of the samples were supplemented with an addition of 80 µg of BTB, corresponding to an ~8-time increase in dye content compared to the quantity of BTB released from the samples after 96 h (Figure 4). The average viability of the cells exposed to the extract of D-TF* from the pH 5 McIlvaine solution was 97±0 % after 1 day, 101±7 % after 3 days, and 90±1 % after 7 days. The extract of D-TF* prepared in a pH 8 McIlvaine solution supported cell viabilities of 100±1 % after 1 day, 93±5 % after 3 days, and 96±3 % after 7 days. The average viability of the cells exposed to the extract of D-FD* from the pH 5 McIlvaine solution was 97±2 % after 1 day, 101±1 % after 3 days, and 85±4 % after 7 days. The extract of D-FD* prepared in a pH 8 McIlvaine solution afforded cell viabilities of 97±2 % after 1 day, 88±7 % after 3 days, and 99±0 % after 7 days.

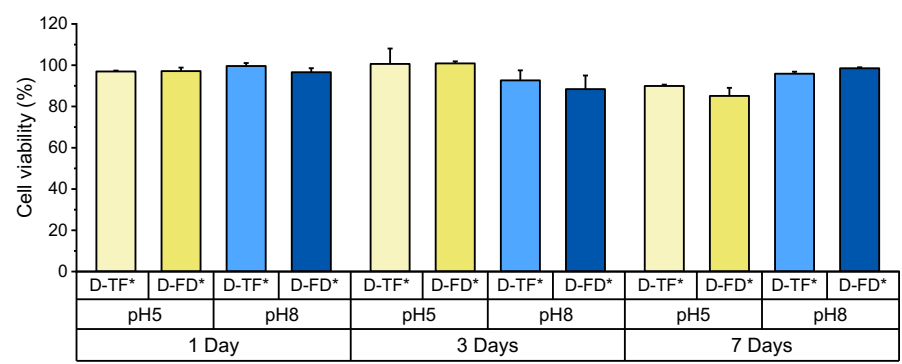

**Figure 8.** Resazurin assay during a 7-day culture of L929 fibroblasts in extracts of D-TF* and D-FD*. Extracts were prepared in McIlvaine solutions adjusted at pH 5 and 8.

The marginal variation in cell viability over the 7-day cell culture period indicates a high tolerability of L929 cells from both drop-cast sample extracts. After 1 day, none of the sample extracts led to statistically different cell viabilities, and the same was true after 3 days. After 7 days, the extracts prepared in the pH 5 McIlvaine solution revealed a statistically significant difference when compared to extracts prepared in the pH 8 McIlvaine solution. The average cell viability of pH 5 extracted samples was 88 %, while the average cell viability of pH 8 extracted samples was 97 %. Although statistically significant reductions in cell viability were measured after 7 days of culture in pH 5 extracts of both D-TF* and D-FD*, all test wells exhibited cell viabilities over 80 %, so that an average cell viability of 92 % was measured. These observations suggest that all sample extracts are compatible with L929 cells and support the suitability of the proposed design to accomplish safe pH monitoring for wound dressing applications.

Further evidence of cytocompatibility is seen in **Figure 9**, which shows the Live/Dead micrographs on day 7 of cell culture. L929 fibroblasts were found to proliferate and displayed a healthy phenotype when exposed to extracts of samples from both acidic and alkaline McIlvaine solutions, further indicating the tolerability of the BTB and any chemical residues extracted from the drop-cast collagen material.

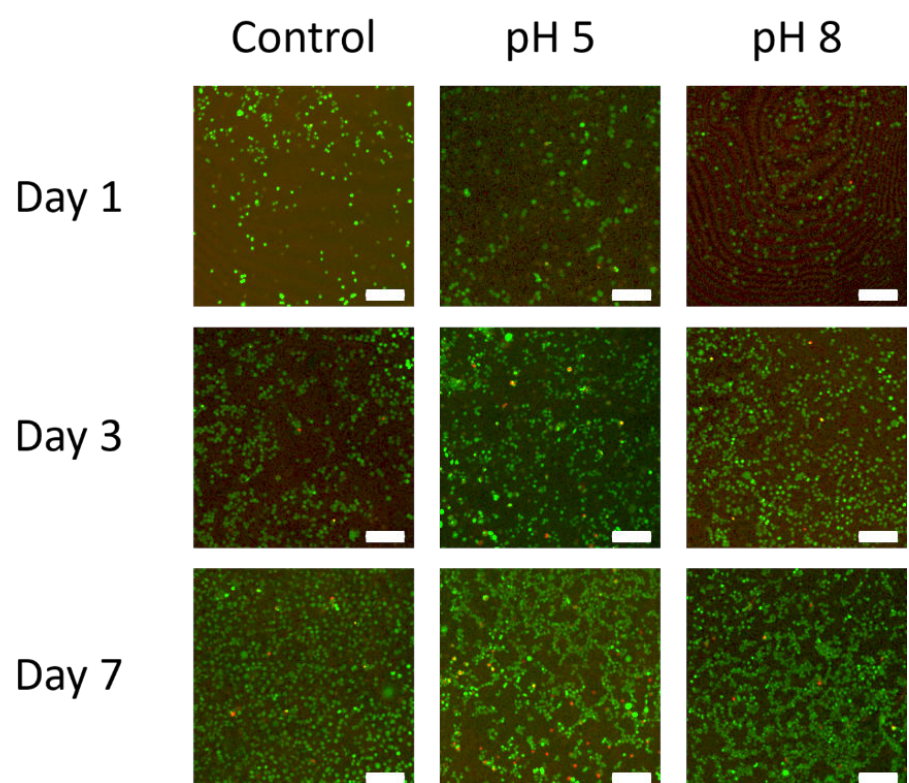

**Figure 9.** Live/dead microscopy images of L929 fibroblasts captured following 1, 3, and 7-day culture on DMEM control, as well as DMEM-containing extracts of TF* prepared in pH 5 and pH 8 McIlvaine solutions. Green: viable cells; red: dead cells. Scale bars: 200 µm.

## 4 Conclusions

A simple wound theranostic dressing prototype was successfully realised by exploiting the colour change capability of bromothymol blue (BTB) at infection-associated alkaline pH, and the development of secondary interactions between BTB and a wound healing collagen network. BTB was integrated into the therapeutic collagen network via either electrospinning or drop-casting, resulting in an average BTB loading efficiency of 99 wt.%. Upon contact with simulated infected wound fluid, both prototypes displayed a visible colour change within one minute. Remarkably, drop-cast samples proved to retain up to 97 wt.% of BTB after 96 hours in simulated wound fluids, equating to an 80 % reduction in dye release (under both simulated healthy and infected wound environments) compared to the electrospun construct bearing BTB-encapsulated PCL fibres. The BTB retention in the drop-cast samples successfully enabled durable infection-sensing capability, as demonstrated by a reversible colour shift following alternating sample incubation in either acidic or alkaline conditions. This surprising dye retention is attributed to electrostatic interactions between the protonated amines of the collagen network and the negatively charged dye molecule, which were not observed in the fast-releasing electrospun construct. The addition of BTB did not affect the morphology and mechanical properties of the collagen hydrogels, while extracts of the drop-cast samples afforded an averaged cell viability of 92 % following a 7-day culture of L929 mouse fibroblasts. The simple, scalable, and cell-/regulatory-friendly design of this wound theranostic offers a new platform for the development of advanced dressings, aiming to support wound healing, reduce hospitalisation time, and enable informed and personalised variations in clinical care. Having demonstrated the infection detection functionality of this collagen-based dressing prototype, the next steps of this research will focus on systematic colour measurements over small pH changes, aiming to develop a reliable calibration curve to quantify and predict infection levels.

**Acknowledgements**

The authors gratefully acknowledge Dr Matthew Hughes, Sarah Myers, and Michael Brookes for technical assistance with circular dichroism, cytotoxicity testing, and SEM, respectively. The Clothworkers' Company (London, UK), the University of Leeds (Leeds, UK) and the Engineering and Physical Sciences Research Council (EPSRC) (Swindon, UK) are also gratefully acknowledged for financial support.

**Supporting Information**

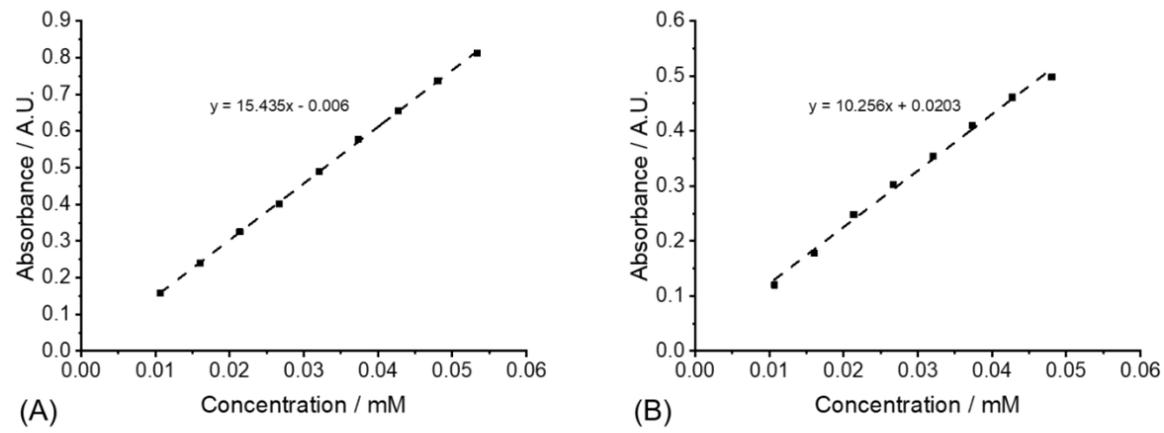

**Figure S1.** Room temperature UV-vis calibration curves for: BTB in ethanol, (A); BTB in acetone, (B).

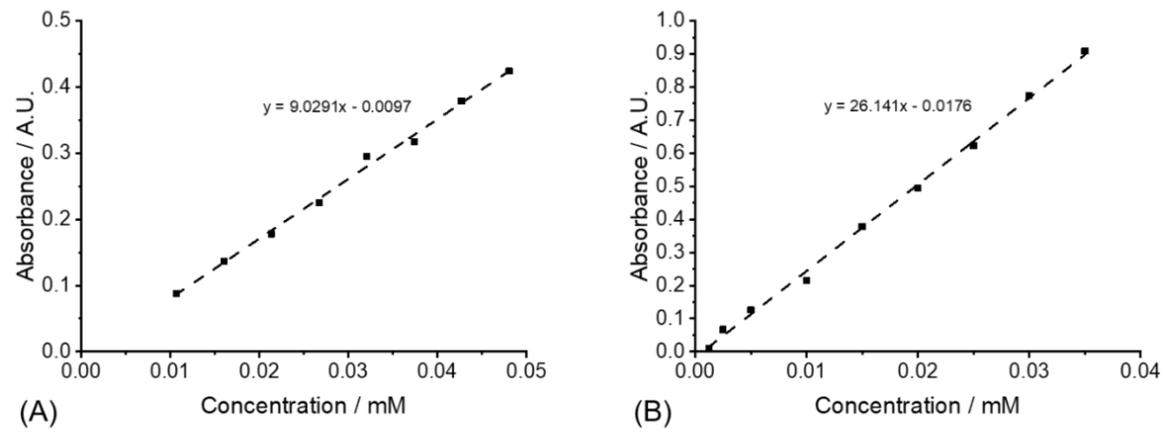

**Figure S2.** Room temperature UV-vis calibration curves for: BTB in pH 5 McIlvaine solution, (A); BTB in pH 5 McIlvaine solution, (B).

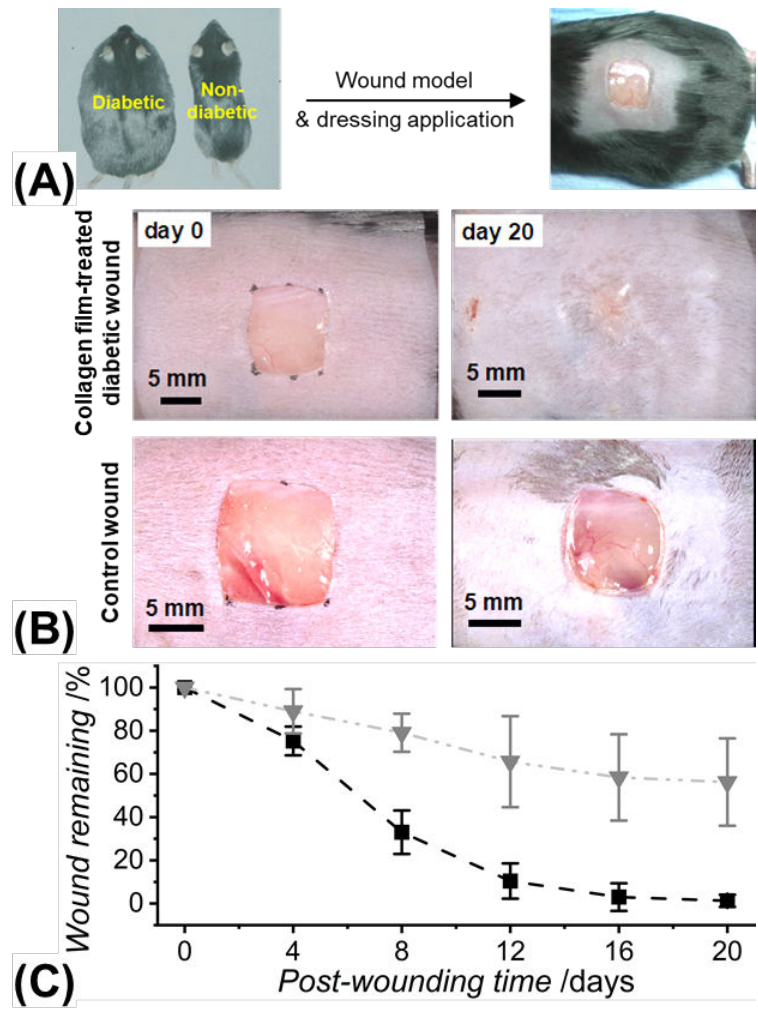

**Figure S3.** Wound healing capability of the collagen-based wound contact layer *in vivo*. (A): Dry films were applied on a full-thickness murine diabetic wound model and replaced every 4 days for an overall period of 20 days. (B): Macrographs captured at day 0 (left-hand image) and day 20 (right-hand image) post-wounding of experimental wounds (top) in receipt of the collagen film or control wounds (bottom) in receipt of a commercial adhesive (Bioclusive™, Systagenix Wound Management). (C): Temporal profiles of the remaining wound area upon receipt of the collagen film (–■–) and commercial adhesive (–··▼–··).

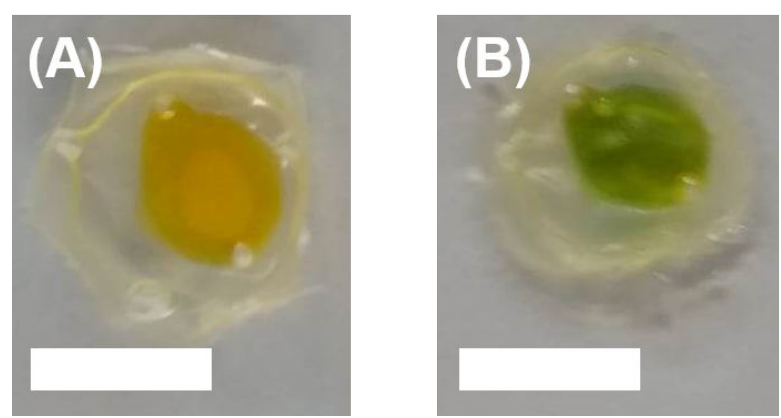

**Figure S4.** pH-induced colour change capability of BTB drop-cast film (prepared with 12-hour drying time) after 1-minute exposure to pH 8 McIlvaine solution. D-TF* before addition of McIlvaine solution (A); D-TF* after 1 minute of exposure to pH 8 McIlvaine solution (B). Scale bars 1 cm.

**Table S1.** Quantification of BTB loading in drop-cast samples (prepared with 12-hour drying time) and electrospun fibres. *LE*: loading efficiency. *L×C×h*: Luminance (*L*), Chroma (*C*) and Hue (*h*) recorded following sample incubation in both pH 5 and pH 8 media.

| Sample ID | *LE* /wt.% | *L×C×h* | |
|---|---|---|---|
| | | pH 5 | pH 8 |
| D-TF*12 | 99±5 | 55×50×76 | 30×11×135 |
| D-FD*12 | 99±5 | 62×64×64 | 44×39×76 |
| D-PCL | 99±4 | 100×0×236 | 100×0×247 |
| D-PMMA-co-MAA | 99±1 | 93×0×127 | 94×0×142 |

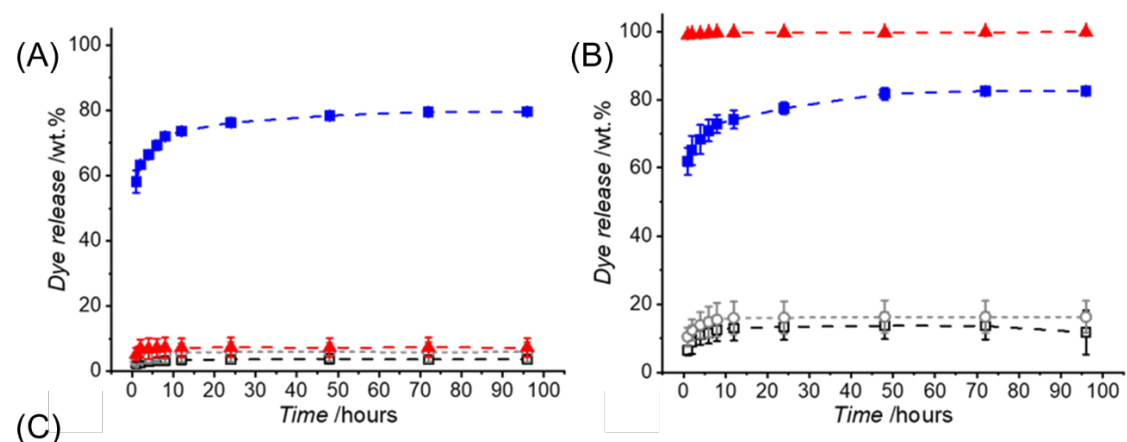

**Figure S5.** Dye release profiles recorded following room temperature incubation in a McIlvaine buffer at pH 5 (A) and pH 8 (B). (–□–): D-TF*12; (--○--): D-FD*12; (–▲–): D-PMMA-co-MAA; (–■–): D-PCL. (C): Quantification of BTB release from drop-cast samples (prepared with 12-hour drying time) and electrospun fibres. $R^*$: fraction released following 96-hour incubation at pH 5 and pH 8. *Zero-order*: coefficient of determination ($R^2$) values obtained via zero-order release profile fitting; *Korsmeyer-Peppas*: $R^2$ values obtained following release profile fitting via Korsmeyer-Peppas model (number in bracket indicates the value of the release exponent).

**Table S2.** Quantification of BTB release from submerged drop-cast samples in simulated healthy (pH 5) wound environments. $R^*$: BTB fraction released after specific time points of sample incubation.

| Sample ID | $R^*$ / wt.% | | | |
|---|---|---|---|---|
| | 1h | 6h | 24h | 96h |
| Sub-D-TF* | 8.2±2.2 | 11.3±3.6 | 14.1±5.7 | 14.8±5.1 |
| Sub-D-TF*12 | 6.5±0.8 | 10.8±0.8 | 13.8±1.5 | 13.8±1.4 |
| Sub-D-FD* | 6.6±1.2 | 11.6±2.0 | 13.3±2.4 | 13.4±2.3 |
| Sub-D-FD*12 | 6.2±2.7 | 11.6±3.9 | 14.1±4.1 | 14.4±4.4 |

**Table S3.** Quantification of BTB release from submerged drop-cast samples in simulated infected (pH 8) wound environments. $R^*$: BTB fraction released after specific time points of sample incubation.

| Sample ID | $R^*$ / wt.% | | | |
|---|---|---|---|---|
| | 1h | 6h | 24h | 96h |
| Sub-D-TF* | 16.6±1.9 | 22.4±3.5 | 24.0±5.0 | 24.0±5.0 |
| Sub-D-TF*12 | 14.3±1.7 | 23.4±5.9 | 24.5±6.7 | 24.5±6.4 |
| Sub-D-FD* | 12.3±3.4 | 18.0±1.9 | 20.0±1.8 | 20.4±1.5 |
| Sub-D-FD*12 | 16.6±4.4 | 21.6±4.9 | 23.1±3.9 | 23.4±3.7 |

## Graphical abstract

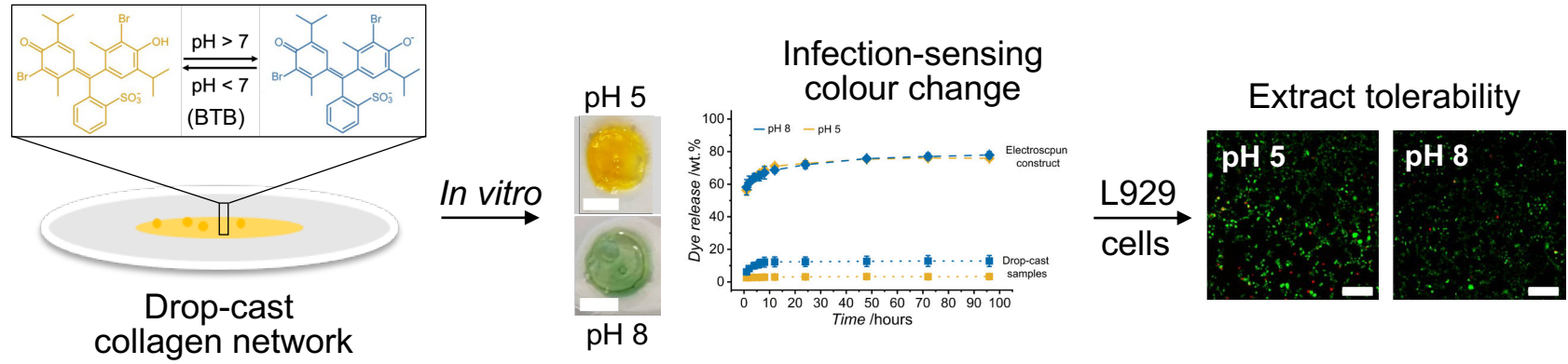